# Large Thermopower with Sign-Alternating Quantum Oscillations in Insulating Monolayer $WTe_2$

Yue Tang[1], Tiancheng Song[1], Haosen Guan[1], Yanyu Jia[1], Guo Yu[1,2], Zhaoyi Joy Zheng[1,2], Ayelet J. Uzan[1], Michael Onyszczak[1], Ratnadwip Singha[3,4], Xin Gui[3], Kenji Watanabe[5], Takashi Taniguchi[6], Robert J. Cava[3], Leslie M. Schoop[3], N. P. Ong[1], Sanfeng Wu[1*]

[1]Department of Physics, Princeton University, Princeton, New Jersey 08544, USA.
[2]Department of Electrical and Computer Engineering, Princeton University, Princeton, New Jersey 08544, USA.
[3]Department of Chemistry, Princeton University, Princeton, New Jersey 08544, USA.
[4]Department of Physics, Indian Institute of Technology Guwahati, Assam 781039, India.
[5]Research Center for Electronic and Optical Materials, National Institute for Materials Science, 1-1 Namiki, Tsukuba 305-0044, Japan.
[6]Research Center for Materials Nanoarchitectonics, National Institute for Materials Science, 1-1 Namiki, Tsukuba 305-0044, Japan.

[*]Correspondence to: sanfengw@princeton.edu

**Abstract**

**The detection of Landau-level-like energy structures near the chemical potential of an insulator is essential to the search for a class of correlated electronic matter hosting charge-neutral fermions and Fermi surfaces, a long-proposed concept that remains elusive experimentally. Here we introduce and demonstrate that the magneto-thermoelectric response of a quantum insulator can reveal critical information not available via other approaches. We report the observation of large thermopower together with quantum oscillations (QOs) in the hole-doped insulating state of monolayer tungsten ditelluride ($WTe_2$) in magnetic fields. The measured low temperature magneto-thermopower exceeds $k_B/e$ by more than an order of magnitude, where $k_B$ is the Boltzmann constant and $e$ the elementary charge. This large thermopower is a characteristic of an insulating state, consistent with high resistivity. However, as the magnetic field is swept, QOs develop in the thermopower, which remarkably undergoes sign-changes that mimic the quantum characteristic of metals due to Landau quantization. The resistivity of the sample is orders of magnitude larger than the value expected from the QOs. The sign-change in the thermoelectric response directly implies the presence of a field-induced Landau-level-like structure at the chemical potential of the insulator. Neither the large thermopower nor the sign-changes can be induced by the metallic gate nearby. Our results demonstrate a new dilemma for investigating low energy excitations in correlated materials featuring mixed quantum characteristics of metals and insulators.**

## Main

Similar to resistivity, thermoelectricity, such as the thermopower measured in the Seebeck effect, probes low energy excitations and transport properties of quantum materials. Conventionally, the thermopower of an electrical insulator is sharply distinguished from that of a metal in both the magnitude and the responses to conditions such as temperature and external magnetic fields. For instance, in metals or semimetals, the thermopower is expected to be less than the natural unit $k_B/e$, ~ 86 μV/K, and diminishes with reducing temperature ($T$) according to the celebrated Mott formula[1]. In contrast, the thermopower of an insulator follows an opposite trend, roughly captured[2,3] by ~ $\Delta/eT$ where $\Delta$ is the activation energy. It can hence be much larger than the natural unit at low temperatures. In magnetic fields, the thermopower of a metal is further sensitive to Landau quantization, revealing the quantum characteristic of a Fermi surface. In this work, we investigate the thermopower in the unusual insulator state of monolayer $WTe_2$, where conventional wisdom is challenged.

While few-layer and bulk $T_d$-$WTe_2$ is a compensated semimetal[4], its monolayer form is an insulator with both strong spin-orbit coupling and large exciton binding. Near charge neutrality, the monolayer exhibits the quantum spin Hall effect and is characterized as a topological excitonic insulator (EI)[5–9] (**Fig. 1a**). When doped with electrons, $WTe_2$ undergoes a superconducting transition[10,11] at a critical carrier density $n_c$ ~ $10^{12}$ $cm^{-2}$, with unusual features[12,13] that lie beyond conventional Landau-Ginzburg description. In sharp contrast, the monolayer, when doped with holes, exhibits a host of unusual properties suggestive of strong correlations at low temperatures ($T$). In typical samples the system remains insulating up to hole densities of ~$10^{13}$ $cm^{-2}$ [6–9,14] (**Fig. 1b**). Chemical potential measurements reveal a sharp peak in the density of state (DOS) located at the valence band edge[9]. This nearly flat dispersion at the valence band edge is accompanied by the formation of the excitonic insulator gap (**Fig. 1a**). Although localization effects are important in 2D systems, they cannot by themselves account for all the observations. The unconventional nature of the hole carriers is also evident in recent experiments on twisted bilayer $WTe_2$, in which a 2D anisotropic Luttinger liquid phase was observed on the hole side and near charge-neutrality, but not on the electron side[15,16]. The nature of hole carriers in 2D $WTe_2$ are so far poorly understood, which is the subject of this study.

## Devices for Detecting Monolayer Thermoelectricity

To examine the Seebeck effect of monolayer $WTe_2$, we employ the device geometry illustrated in **Fig. 1c**. A heater, made of a thin, narrow and curved Pd strip, is fabricated next to the monolayer $WTe_2$, which is encapsulated between two graphite/hexagonal boron nitride (hBN) stacks (details in **Extended Data Fig. 1**). Metal electrodes are fabricated to extract electrical signals generated in the monolayer. The vdW stacks allow us to apply gate voltages to either the top ($V_{tg}$) or bottom ($V_{bg}$) graphite layers, as well as a temperature gradient $-\nabla T$ within the monolayer plane using the Pd strip heater. An optical image of a device (D1) is shown in **Fig. 1d**. In addition to the thermopower, we also carried out electrical measurements on the device. **Figure 1e** displays the four-probe conductance ($G_{xx}$) as a function of gate-induced carrier density $n_g$, at selected $T$. The residual conductance in the insulator state due to transport in the helical edge mode is rapidly suppressed in strong magnetic fields ($B$) (**Fig. 1f**). Note that here the measurement involves long (> 10 μm) and hence resistive edge channels (see the inset of **Fig. 1f**). The monolayer is insulating with a rapidly suppressed conductivity at low $T$ over the entire hole-doped regime (**Fig. 1e**). By contrast, on the electron-rich side we observe signatures of metal-insulator transition and high conductance. These observations are consistent with previous reports[6–8,14]. A similar device geometry has recently enabled successful measurements of the vortex Nernst effect in the electron-rich superconducting state of the monolayer[12,13]. Here we focus on detecting the thermopower of monolayer $WTe_2$ on the insulating hole-rich side.

## Sign Alternating Thermopower and Quantum Oscillations

For non-interacting fermions, the thermoelectric response function $\alpha = J/|\nabla T|$ ($J$ is the charge current density) is given by $\alpha = 2e\int dE\left(-\frac{\partial f_0}{\partial E}\right)\frac{(E-\mu)}{k_B T}\mathcal{N}(E)v^2\tau$ , where $f_0$ the Fermi-Dirac distribution, $\mu$ the chemical potential, $\mathcal{N}(E)$ the DOS, $v$ the group velocity and $\tau$ the transport lifetime (the thermopower is $S = \alpha/\sigma$) [1]. From the kernel in the equation, we see that $\alpha$ is negative (particle-like) if the derivative $(\partial\mathcal{N}/\partial E)_\mu > 0$, and positive (hole-like) if $(\partial\mathcal{N}/\partial E)_\mu < 0$. In a magnetic field $B$, $\mathcal{N}(E,B)$ is the sum of the DOS of individual Landau levels (LLs), approximated by a Lorentzian centered at $E_n = n\hbar\omega_c$ ($\omega_c$ is the cyclotron frequency, $\hbar$ is the plank constant and $n$ is an integer index). We infer that, as $B$ is varied, the longitudinal thermoelectric response function $\alpha_{xx}(B)$ changes sign each time $E_n$ crosses $\mu$ (**Fig. 2a**). As discussed below, the periodic changes in sign provide a crucial test of the intrinsic nature of the LLs. Such sign changes of $\alpha$ due to Landau quantization have been observed in semimetals including graphene[17,18], few-layer $WTe_2$ [19], and quantum wells[20]. It also occurs due to formation of discrete levels in quantum dots[21].

We investigate the magneto-response of monolayer $WTe_2$. **Figure 2b** plots the resistance $R_{xx}$ of the monolayer (device D1) at the gate voltages ($V_{tg}$ = -3.9 V; $V_{bg}$ = - 4.5 V). As $T$ decreases, $R_{xx}$ diverges as characteristic of an insulator, reaching ~ 700 KΩ at 2 K. **Figure 2c** plots the magnetoresistance curves which display Shubnikov–de Haas (sdH)-like oscillations. This is surprising as the sample is far from being metallic. The observed QOs in resistance is consistent with the previous report[14].

Our key finding is the QOs in the thermopower in the insulating state. A weak temperature gradient $-\nabla T$ is established in the monolayer by the heater. In the $I_s$ configuration (**Fig. 2d**), the gradient $-\nabla T$ drives a thermoelectric current $I_S$ that flows between the pair of electrodes shorted together (all other contacts are floated). With increasing $B$, $I_S$ develops QOs, consistent with the oscillations in $R_{xx}$. Importantly, $I_S$ changes sign whenever $R_{xx}$ displays a local peak. Likewise, oscillations are also observed in the $V_s$ configuration (**Fig. 2e**), in which the selected electrodes are now left open. The voltage difference $V_S$ across them oscillates *vs.* $B$, as in the trace of $I_S$. The oscillations are periodic in $1/B$ and a fast Fourier transformation (FFT) yields the same period of ~ 55.8 T for all three curves (**Fig. 2f**). In large $B$, the curve of $V_S$ reveals additional sign-changes that coincide with weak maxima seen between the main peaks in $R_{xx}$. Compared with $I_s$, $V_S$ seems more sensitive to these features which may arise from, e.g., spin splitting.

We note that the amplitude of $V_S$ in the high $B$ regime is on the order of ~ μV whereas $I_S$ is ~ pA. Hence the value of $V_S/I_S$ matches the resistance of monolayer $WTe_2$ qualitatively, a proof that the thermoelectricity is indeed generated in the $WTe_2$ layer. At low fields ($B$ < 2 T), $V_S$ typically vanishes in our devices while $I_S$ might be of a finite value. While we don't fully understand the low-$B$ behavior, we suspect that the presence of topological edge mode, as well as disorders and localization effects, may play a role in this regime. The edge mode is, however, largely suppressed in high $B$ and would not be able to generate any signals that are periodic in $1/B$. It is worth mentioning that in graphene the low field thermoelectricity data is less understood as well[17,18].

**Gate and Temperature Dependence**

The sign-alternating thermoelectric QOs are observed over a wide hole doping regime. **Fig. 3a** plot $R_{xx}$ *v.s.* $V_{tg}$ while fixing $V_{bg}$, taken at various $T$, again demonstrating the insulating behavior in resistance. On the wide hole-rich side, the thermoelectric signals ($V_S$, as shown in **Fig. 3b**) display QOs with a frequency that is tunable by $V_{tg}$. The oscillation frequency is however independent of $V_{bg}$ even though $V_{bg}$ does change the overall doping of the sample as indicated in the resistance data (**Figs. 3c** & **d**). The single-gate dependence of the QO frequency is consistent with previous transport results in the same hole-doped regime[14,22]. We note that while the $V_{tg}$-dependent QO branch is dominant on the hole side here, there could be other branches not seen in the measurements. The observation that the QO frequency of one branch in a multi-carrier system tends to zero doesn't imply zero total carrier density. Note that on the electron side, three QO branches were observed previously in resistance (Extended Data Fig. 4 in ref [14]). The key observation of alternating sign changes induced by $B$ in the $V_S$, like that seen in **Fig. 2**, is true over a wide gate range (**Figs. 3e-i).** The phenomena hence cannot be attributed to any singularity that is specific to critical carrier density or features like van Hove singularity. The sign-changes in the thermoelectric QOs and the gate-dependent behaviors are highly reproducible. In **Figs. 3j-n**, we present consistent data obtained from another device (D2) (more details in **Extended Data Figs. 2-5**). **Fig. 4** presents the $T$ dependence of $V_S$ and the QOs. Upon warming up the device to ~ 5 K, the QOs are suppressed altogether along with the vanishing of $V_S$.

The presence of a QO branch with a single-gate dependent frequency was previously argued as an evidence for attributing the oscillations to the graphite gate, a scenario discussed in the initial report of the QOs[14] and later further proposed in ref[22]. However, this argument relies on an unjustified assumption that the atomically thin $WTe_2$ insulator screens electric fields perfectly as in the case of an excellent metal [23,24]. The same observation could therefore be argued as evidence against the graphite gate scenario. Other issues raised by this interpretation have been discussed elsewhere[23,24]. In the next section, we will discuss key observations in the thermopower (amplitude and sign-changes) revealed in this work that directly rule out the possibility of attributing the QOs to the graphite gate. The single-gate dependence of the QO frequency is indeed a puzzling observation in $WTe_2$ and a comprehensive understanding is still needed. However, we note that in systems with multi-band carriers, especially with excitons, the expectation for the gate-dependence in a dual-gated device is no longer simple and universal. It is possible that the presence of excitons can promote the single gate dependence of certain QO branches (see a proposed scenario in Extended Data Fig. 4 in ref [14]).

## Large Thermopower and Exclusion of the Graphite-Gate Scenario

The existence of QOs in an insulating material is anomalous. Clearly, we should exclude artefactual mechanisms that could mimic the QOs. In the graphite-gate proposal[22], the carrier density $n$ of $WTe_2$ undergoes tiny changes, $\delta n$, due to small modulations of the gate capacitance in response to QOs in the graphite layer. The resistance of the insulating $WTe_2$ layer is regarded as a detector of the graphite QOs. An oscillating $\delta n$ may produce changes in $R_{xx}$. However, it can never change the character of the carriers from being "electron-like" to "hole-like" induced by $B$. This is however key information revealed by the sign-alternating thermopower in $WTe_2$, which is true regardless of $n_g$ on the (substantially) hole doped insulating side. The thermoelectric current $I_s$ can never change sign since $\delta n < n$ (the same argument applies to $V_s$). The present results, with sign-alternating thermopower observed over a wide hole doping range, exclude all mechanisms based on artefactual modulations of the itinerant carrier density of monolayer $WTe_2$. It thus precludes such artefacts arising from the graphite gate. We also emphasize that our measurements detect the thermopower signals in the DC limit (**Extended Data Fig. 6**), hence any capacitance related coupling effects are not relevant.

Moreover, the observed thermopower is large and can only be attributed to an insulator. In **Extended Data Fig. 7**, we estimate the temperature difference to be ~ 14.7 mK between the two neighboring contacts when a heater power of 250 nW is applied. For data shown in **Fig. 2**, we only applied ~ 16.9 nW heater power, suggesting that a temperature difference on the order of ~ 1 mK produces a $V_s$ up to 4 μV (**Fig. 2e & Extended Data Fig. 8**). In **Extended Data Fig. 9**, we present finite element simulation for the temperature distribution, which yields a consistent result (see Methods). This yields a thermopower on the order of 4,000 uV/K, which exceeds $k_B/e$ by nearly 50 times, a clear characteristic of an insulator. The only possible source of such a large thermopower in our device is the insulating $WTe_2$. The graphite gate, as a semimetal can only produce a thermopower that is less than $k_B/e$. Indeed, we have performed thermopower measurements on graphite of a similar thickness compared to the graphite gate, no thermopower signal is detectable at all using the same low heater power (**Extended Data Fig. 10**). To observe a finite thermpower signal in the graphite, we need to apply a heater power that is about 350 time higher (**Extended Data Fig. 11**), consistent with expectation of low thermopower due to its metallic nature. Under such a high temperature gradient, the graphite thermpower is indeed observed, allowing us to reveal the true QOs of graphite. They are clearly distinguished from that of the $WTe_2$. All these observations straightforwardly rule out the graphite gate scenario. We conclude that the large thermopower with sign alternating QOs observed in the $WTe_2$ devices is an intrinsic property of the $WTe_2$ layer.

## The Dilemma and Interpretations

In addition to $WTe_2$, QOs in insulators, with a periodicity in $1/B$, have been reported in several systems, including Kondo insulators[25–28] and α-$RuCl_3$[29]. A comprehensive understanding of the oscillations in these cases is still lacking and the interpretations are open including both intrinsic and extrinsic origins[22,27,30–45]. $WTe_2$ is distinct as its parent state is a topological excitonic insulator and the mechanisms are likely different in different materials (see more discussions in ref[46]). In addition to the graphite gate scenario, another mechanism proposed for $WTe_2$ regards magnetic-field induced gap modulations in the excitonic insulator state[37,38]. This is a proposal for the charge neutrality where the excitonic gap forms[37,38], only valid when the number of electrons and holes are equal.[8,9] The theory is not applicable to the wide hole-doped regime, the focus of this work.

Our results here introduce thermopower as a powerful approach in revealing critical information not available using other methods and establish the observation of QOs in monolayer $WTe_2$ in two distinct measurables (resistivity ϱ and thermopower $S$). The observed values of both ϱ and S are so large that the monolayer is clearly in an electrically insulating state, yet dramatic QOs, the quantum characteristic of metal, develop in both quantities above an onset field $B_{onset}$~ 3 T (**Fig. 2d**). This suggests a quantum mobility of quasiparticles $\mu \sim 1/B_{onset} \sim 3300$ cm$^2$/(Vs). In the standard analysis for SdH oscillations, the period of QOs (~ 55.8 T, **Fig. 2f**) indicates a nominal carrier density $n \sim 2.7 \times 10^{12}$ cm$^{-2}$ (assuming spin degeneracy). Consequently, we expect from the QO data a conductivity $\sigma = ne\mu \sim 1400$ μS, and hence a good metallicity in the sample. In sharp contrast, the resistance (**Figs. 2b & c**) yields a strongly insulating state with a conductivity that is ~3 orders of magnitude lower. This is the essential dilemma in the Landau quantization problem of monolayer $WTe_2$.

The striking constraints inferred from the new observations in thermopower invite us to examine the neutral-fermion scenario[14]. The sign-changing thermoelectric QOs imply that $\mathcal{N}(\mu)$ is strongly modulated in $B$, much like conventional LLs. The essential dilemma - the violation of $\sigma = ne\mu$ - can be reconciled if we assume the high-mobility quasiparticles are charge neutral, but still responsible for the observed QOs. One scenario is that holes in monolayer $WTe_2$ undergo spin-charge separation at low *T*, i.e., holes split into charged holons and charge-neutral spinons. The experimental observations of a high DOS at valence band edge of the monolayer[9] and the appearance of a 2D anisotropic Luttinger liquid phase on the hole side of twisted bilayers[15,16] indicates that spin-charge separation in $WTe_2$ could be possible. If these assumptions are valid, then the localization of holons can cause insulating behavior in the hole-doped regime of the monolayer whereas the mobile spinons form neutral Fermi surfaces. In *B*, the spinon band undergoes Landau quantization, similar to that proposed in the context of quantum spin liquid[47] and mixed-valence insulators[43,48]. Both electrical and thermoelectric transport can detect the QOs induced by spinon Landau quantization according to the Ioffe–Larkin rule[49,50]. The spinon LLs are responsible for the sign-changes of the thermoelectric signals. Our results invite an in-depth analysis of hole carriers in $WTe_2$ and thermoelectric transport in spin-charge separated 2D systems.

## Summary and Outlook

The observation of QOs in both electrical and thermoelectric transport places $WTe_2$ at a unique position for investigating unconventional Landau quantization and possible neutral fermions in insulators. The essential dilemma is that the resistivity of the sample is orders of magnitude larger than what would conventionally be expected from the QO data. The detection of the same underlying physics using distinct experimental approaches is key to fully addressing the paradox of QOs in insulators. Beyond transport, recent developments of far-infrared magnetooptical spectroscopy for 2D materials at millikelvin temperatures[51] may provide further insights. Beyond $WTe_2$, the results and approaches developed along this endeavor will be useful in general for discovering next-generation quantum phenomena in correlated 2D crystals and vdW stacks, especially insulators with fractionalized charge-neutral excitations[46].

## Materials and Methods

### Device fabrication

The procedure of device fabrication is similar to that described in our previous works[8,12,14]. (*i*)We first fabricated a bottom van der Waals (vdW) stack consisting of hexagaonal boron nitride (hBN) and graphite flakes, followed by depositing a thin layer of metal electrodes and the heaters on the stack. The hBN and graphite flakes were mechanically exfoliated onto 285 nm $SiO_2$/Si substrates and selected after the examinations under optical and atomic force microscopes (AFM). The standard 2D dry transfer technique was employed for creating the bottom graphite/hBN stack on a $SiO_2$/Si (insulating) substrate with prepatterned Ti/Au metal pads for wire bonding as well as alignment markers. The thin metal electrodes and microheaters (2 nm Ti/10 nm Pd) were deposited on top of the bottom vdW stacks, followed by another deposition of 5 nm Ti/55 nm Au that connects the thin electrodes on the vdW stack to the prepatterned thick metal wire-bond pads. All metals are patterned and deposited using the standard e-beam lithography and metal deposition tools in a clean room. (*ii*) the prepatterned bottom stacks are then cleaned and examined carefully under AFM to confirm its cleanness with an atomic resolution. The AFM tip cleaning procedure using the contact mode is typically applied to make sure the best quality of the surface. (*iii*) Another key part of the device is a high-quality top stack consisting of graphite/hBN/monolayer $WTe_2$. High-quality $WTe_2$ bulk crystals were exfoliated onto 285 nm $SiO_2$/Si substrates in an argon-filled glovebox (with water and oxygen concentrations less than 0.1 ppm). Monolayer $WTe_2$ flakes were identified by the standard optical contrast that was routinely applied. The same dry transfer technique was employed to make the top hBN/graphite/$WTe_2$ stack, which was released onto the pre-patterned bottom stack in the glovebox. The optical and AFM images at selected stages of the process and more details about the device structures can be found in **Extended Data Fig. 1**.

### Electrical transport measurements

The electrical transport measurement was conducted in a Quantum Design Dynacool system (with a base temperature of ~ 1.8 K), equipped with a superconducting magnet (up to 9 T). Four-probe resistance measurements were performed using the standard ac lock-in technique with a low frequency ~ 7 Hz. A small voltage excitation of 2 mV typically is applied to the source electrodes while the current flow through the drain is measured with a current preamplifier (DL Instrument 1211). The voltage drop between two longitudinal probes are recorded with a low noise voltage preamplifier (DL Instrument 1201).

### Thermoelectric measurements

In the thermoelectric measurement, a current is applied to a microheater (of resistance ~ 10 KΩ) fabricated next to the monolayer $WTe_2$ to generate a temperature gradient. As illustrated in Fig. 1C, the microheater is made of a thin, narrow and curved metal stripe (~ 200 nm wide and ~ 12 nm thick). During measurements (configurations shown in Figs. 2d & e), an alternating current was applied to a microheater at low frequency of ω ~ 4 Hz, which produces an alternating temperature gradient oscillating at the frequency of 2ω. The lock-in technique is employed to extract thermoelectric signals ($V_S$ or $I_S$) at the frequency of 2ω, like the standard electrical transport measurements. The same device structure and measurement technique has been applied successfully to detect the vortex Nernst signals[12] of monolayer $WTe_2$. A Keysight 33500B series waveform generator supplies the current to the heater, and the thermoelectric signal generated at $WTe_2$ contacts is fed into a DL Instruments 1211 current preamplifier (in the $I_S$ configuration) or DL Instrument 1201 voltage amplifier (in the $V_S$ configuration) before an SR830 lock-in amplifier. Detailed measurement scheme can be found in **Extended Data Fig. 3**.

The effect of heater power is shown in **Extended Data Fig. 8**. In the DC measurement (**Extended Data Fig. 6**), no lock-in amplified is used. A DC voltage source is used to apply a static current to the heater and DC voltages between selected contact are recorded as a function of time, after a voltage preamplifier (DL Instrument 1201).

**Finite element simulation on the temperature gradient**

The direction of the temperature gradient is determined by device geometry such as the location of the heater, electrodes and the whole vdW stack structures, which are fixed once the device is fabricated. The software COMSOL is employed to simulate the stationary heat distribution of device D1 under thermoelectric measurement. **Extended Data Fig. 9a** presents the top view of the corresponding computer model of the full material stack, which considers the realistic material composition and thickness different layers of the device. There are three thermal components, including the heat source that is assigned to the center of the microheater with constant heat power $P$, the cold source which is served by outer thick metal electrodes with a constant temperature $T_0 = 2$ K), and the heat transfer within the sample and between sample and the environment with a coefficient $h$. The thermal conductivities of graphite, BN, and Pd are assigned as 1, 0.01, and 72 W/(m·K), respectively, while their heat capacities are 0.01, 0.4, and 240 J/(kg·K), the densities are 2260, 2100, and 12000 kg/m$^3$, respectively. These values at 2 K refer to refs[53–57]. The coefficient $h$ is set to be 3 W/(m$^2$·K) in the simulation. **Extended Data Fig. 9b** presents the zoomed-in thermal distribution under $P$ = 16.9 nW.. The simulated temperature gradient of the two contacts measured in the main text is 0.8 mK, consistent with the estimate based on direct local resistance measurements in **Extended Data Fig. 7**.

**Acknowledgments**

This work is mainly supported by the NSF through the Materials Research Science and Engineering Center (MRSEC) program of the National Science Foundation (DMR-2011750) through support to S.W., L. M. S. and R.J.C., and a CAREER award (DMR-1942942) to S.W. S.W. also acknowledges support from AFOSR through a Young Investigator Award (FA9550-23-1-0140), ONR through a Young Investigator Award (N00014-21-1-2804), the Gordon and Betty Moore Foundation through Grant GBMF11946 and the Sloan Foundation. L.M.S. acknowledges support from the Gordon and Betty Moore Foundation's EPIQS initiative through Grant GBMF9064 and the David and Lucile Packard Foundation. S.W. and L.M.S. acknowledge support from the Eric and Wendy Schmidt Transformative Technology Fund at Princeton. R. J. C. at Princeton acknowledges support from the Gordon and Betty Moore Foundation through grant GBMF-9066 and the US DOE division of Basic Energy Sciences (DE-FG02-98R45706). T.S. acknowledges support from the Princeton Physics Dicke Fellowship program. A.J.U acknowledges support from the Rothschild Foundation and the Zuckerman Foundation. K.W. and T.T. acknowledge support from the JSPS KAKENHI (Grant Numbers 20H00354 and 23H02052) and World Premier International Research Center Initiative (WPI), MEXT, Japan.

**Author contributions**

S.W. conceived, designed and supervised the project. Y.T. fabricated the devices and performed measurements, assisted by T.S., H.G., Y.J., G.Y. Z.J.Z., A.J.U., and M.O. R.S. L.M.S., X.G. and R.J.C grew bulk $WTe_2$ crystals. K.W. and T.T. provided hBN crystals. S.W., N.P.O. and Y. T. analyzed the data, interpreted the results, and wrote the paper with input from all authors.

**Competing interests**

Authors declare that they have no competing interests.

**Data availability**

All data needed to evaluate the conclusions are presented in the paper. Additional data related to this paper are available from the corresponding author upon reasonable request.

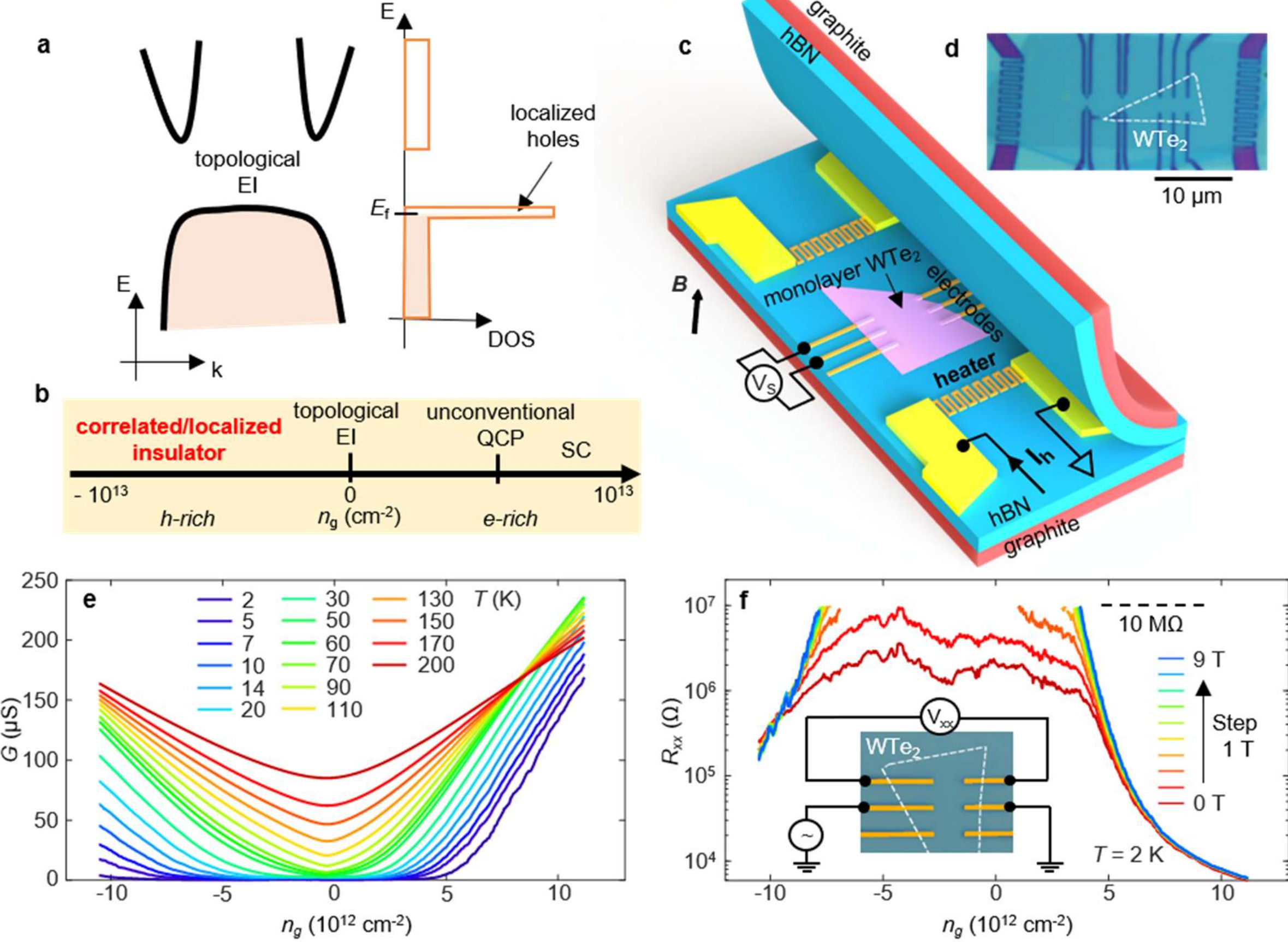

**Fig. 1. The hole-doped insulator of monolayer $WTe_2$ and the device design.** (**a**) Left: a schematic of the electronic band structures of monolayer $WTe_2$, which features a topological excitonic insulator gap at charge neutrality. Right: a qualitative illustration of the DOS, which features a sharp peak near the valence band top. (**b**) A brief summary of the electronic phases observed in monolayer $WTe_2$, tuned by the gate-induced carrier density $n_g$. (**c**) A cartoon illustration of the device geometry employed in this study. (**d**) An optical microscope image of a final device (D1), where the monolayer $WTe_2$ is outlined by the white dashed line. (**e**) Four probe conductance $G_{xx}$ of monolayer $WTe_2$ (D1) as a function of $n_g$, taken at various $T$ as labeled. (**f**) $n_g$-tuned four probe resistance ($R_{xx} \equiv 1/G_{xx}$) taken in various magnetic fields ($B$) applied in direction normal to the 2D flake, showing the rapid suppression of the edge mode conductance. The inset shows the transport measurement configuration.

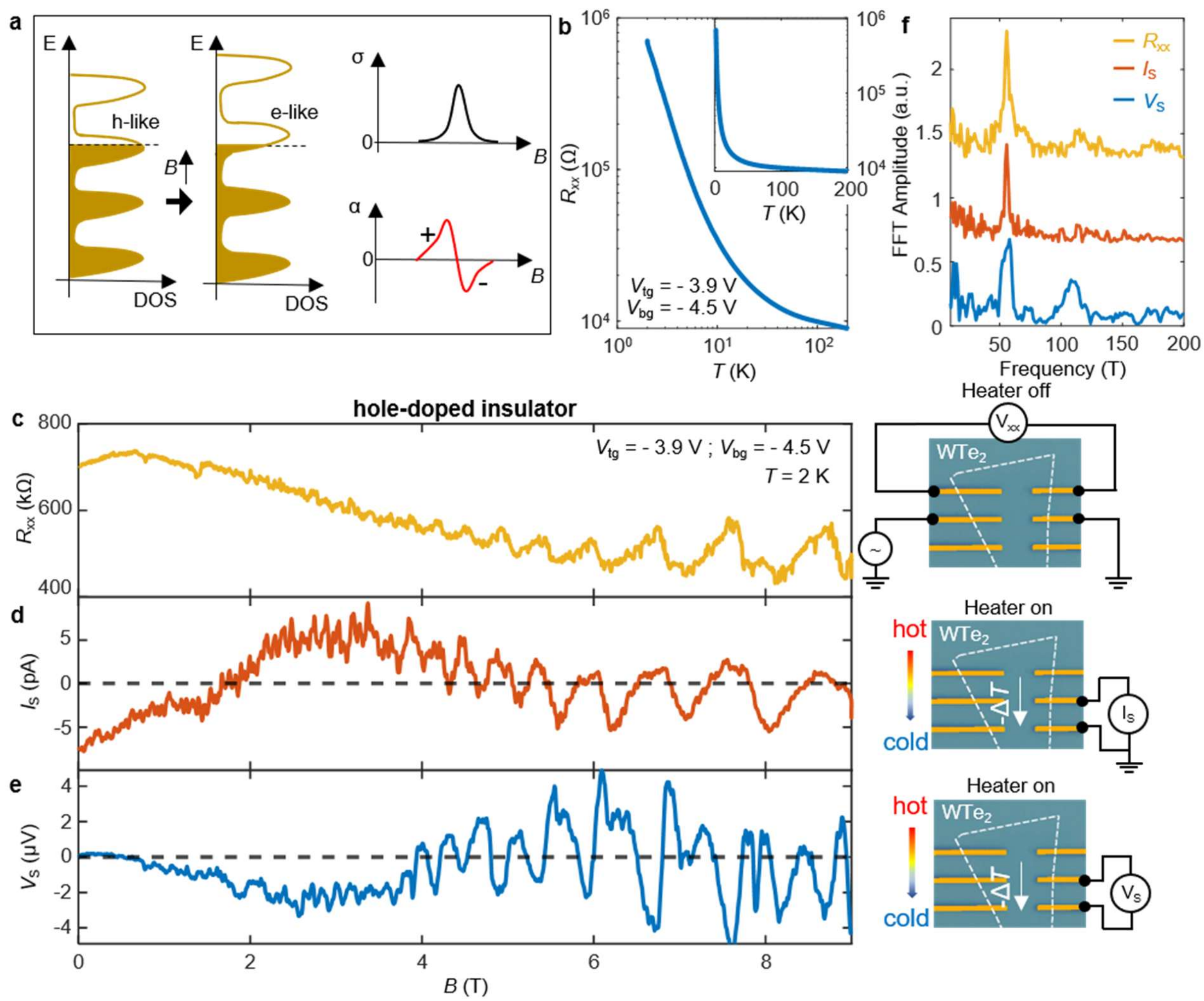

**Fig. 2. Sign-change thermoelectric QOs of the hole-doped insulating monolayer $WTe_2$.** (**a**) Cartoon illustration of Landau quantization in 2D electron gas and its manifestation in conductivity and thermoelectric response with varying $B$. Dashed lines indicate the Fermi energy. The thermoelectric response function α voltage develops alternating sign-changes when $B$ increases. (**b**) $T$-dependent $R_{xx}$ of D1 taken at $V_{tg}$ = - 3.9 V and $V_{bg}$ = - 4.5 V, displaying the insulating behavior. Inset: the same resistance data plotted in a linear scale. (**c**) Magneto-resistance of monolayer $WTe_2$ at the same gate voltages, taken at 2 K. The measurement configuration is shown on the right. (**d**) The thermoelectric current response of the same state, with the measurement configuration shown on the right. (**e**) The Seebeck voltage of the same state, with measurement configurations shown on the right. The dashed horizontal lines indicate the zero value. The heater power is set as 16.9 nW for **d** and **e**. (**f**) Normalized FFT of the QO data shown in (**c**-**e**), with matched colors. In addition to the main frequency, we also observe its weak second harmonic peak. At high $B$, the signals undergo finer splitting (**e**), which may arise from splitting of the Landau levels.

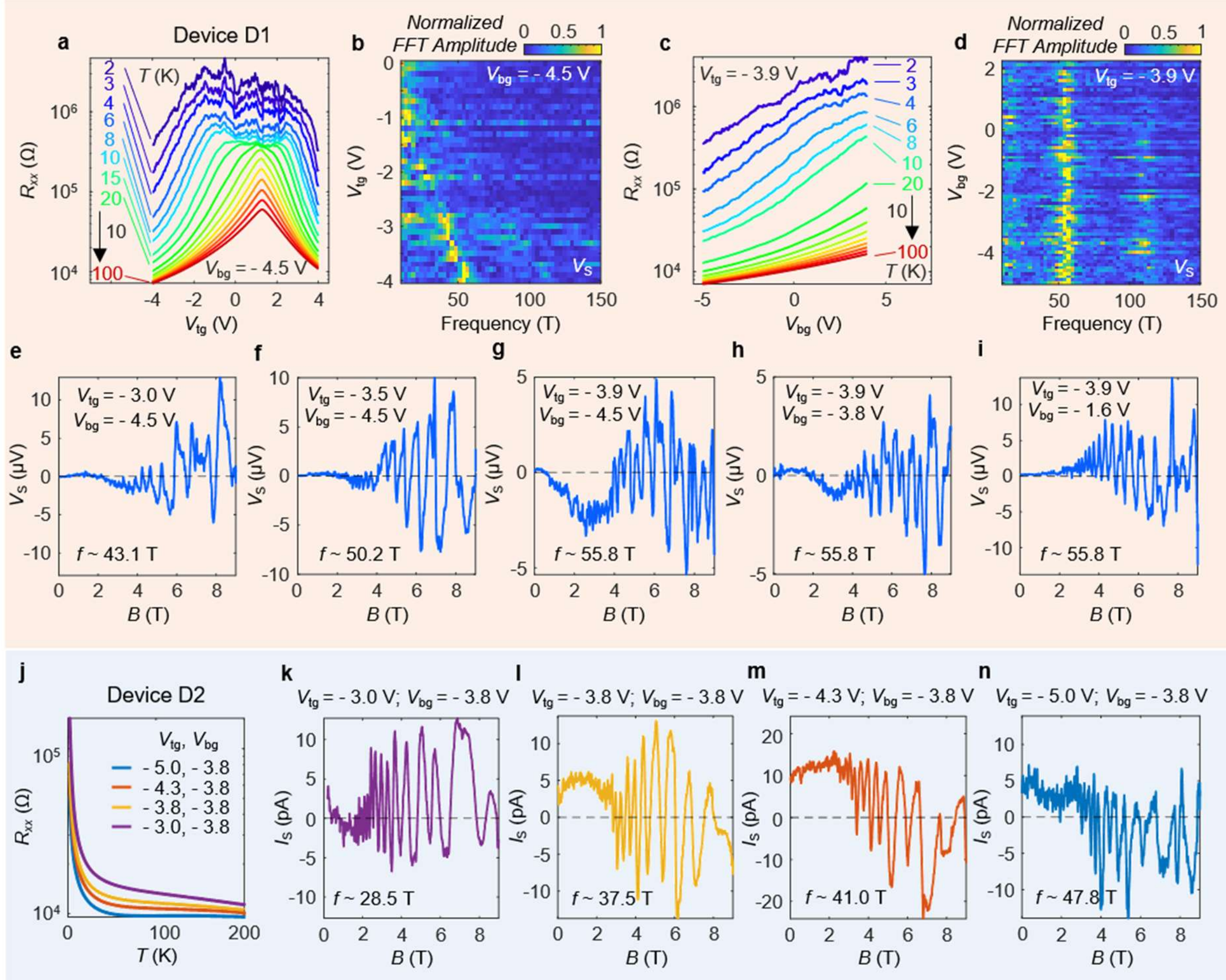

**Fig. 3. Gate dependence of the thermoelectric QOs.** (**a**) $R_{xx}$ as a function of $V_{tg}$ at a fixed $V_{bg}$, taken at various $T$ as indicated by the color. (**b**) A map of normalized FFT amplitude of QOs observed in $V_S$ under varying $V_{tg}$. $V_{bg}$ = - 4.5 V. (**c**) $R_{xx}$ as a function of $V_{bg}$ at the fixed $V_{tg}$, taken at various $T$ as indicated by the color. (**d**) A map of normalized FFT amplitude of QOs observed in $V_S$ under varying $V_{bg}$. $V_{tg}$ = - 3.9 V. (**e-i**) The Seebeck voltage $V_S$ as a function of $B$ taken under various gate configurations as indicated. The QO frequency $f$ from FFT analysis is indicated in each panel. Dashed lines indicate the zero value. Data shown in (**a-i**) are taken from device D1. (**j**) $T$-dependent $R_{xx}$ curves taken at selected gate configurations on the hole-rich side of device D2. (**k-n**) The thermoelectric current $I_S$ observed in this device as a function of $B$, taken at the corresponding gate configurations shown in **j**, respectively. The corresponding QO frequencies are shown in each panel. Measurement configurations for device D2 are shown in **Extended Data Fig. 3**. $T$ = 2 K if not otherwise specified.

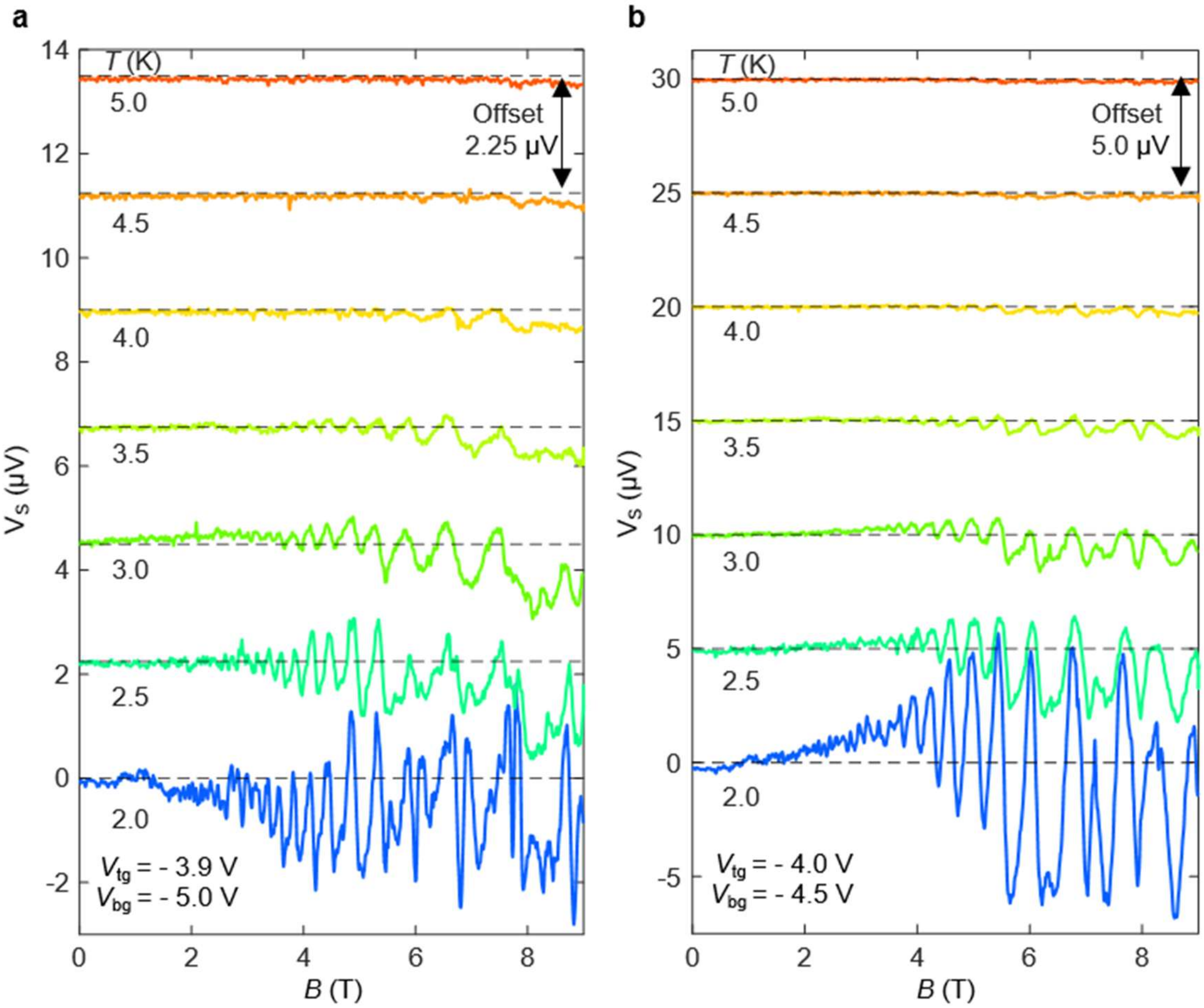

**Fig. 4. Temperature dependence of the QOs. (a)** $V_S$ as a function of $B$, taken at $V_{tg}$ = - 3.9 V and $V_{bg}$ = - 5.0 V, at various $T$ as indicated. **(b)** The same measurement but taken at $V_{tg}$ = - 4.0 V and $V_{bg}$ = - 4.5 V. Data are taken from device D1. The curves are offset by a constant in y axis for a better visualization and the zero value of $V_S$ for each curve are indicated as the dashed line.

## Extended Data Figures

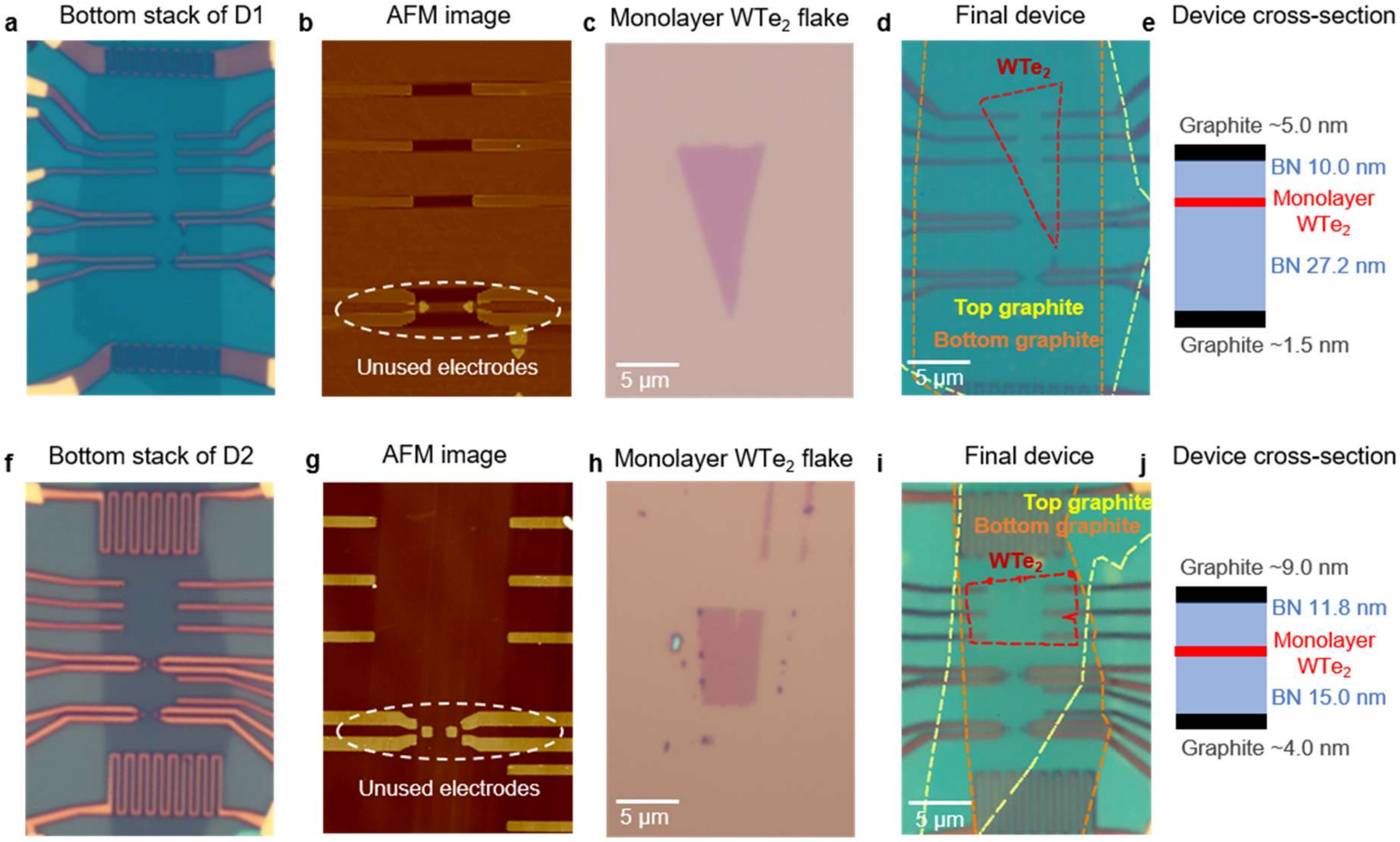

**Extended Data Fig. 1. Details of device fabrication.** (**a**) Bottom graphite/hBN stack and metal electrodes of device D1. (**b**) an AFM image of the prepatterned bottom stack shown in **a**. (**c**) An optical microscope image of monolayer $WTe_2$ flake used for D1. (**d**) An optical microscope image of the final device D1. The red dashed line highlights the monolayer $WTe_2$ flake while yellow and orange dashed lines outline the top and bottom graphite layers, respectively. (**e**) Schematic of the cross section of device D1, with the thickness of each layers indicated. (**f-j**) The same description of device D2, with an optical image of the bottom stack (**f**), an AFM image (**g**), the monolayer $WTe_2$ flake (**h**), an optical image of the final device (**i**), and the cross-section schematic (**j**).

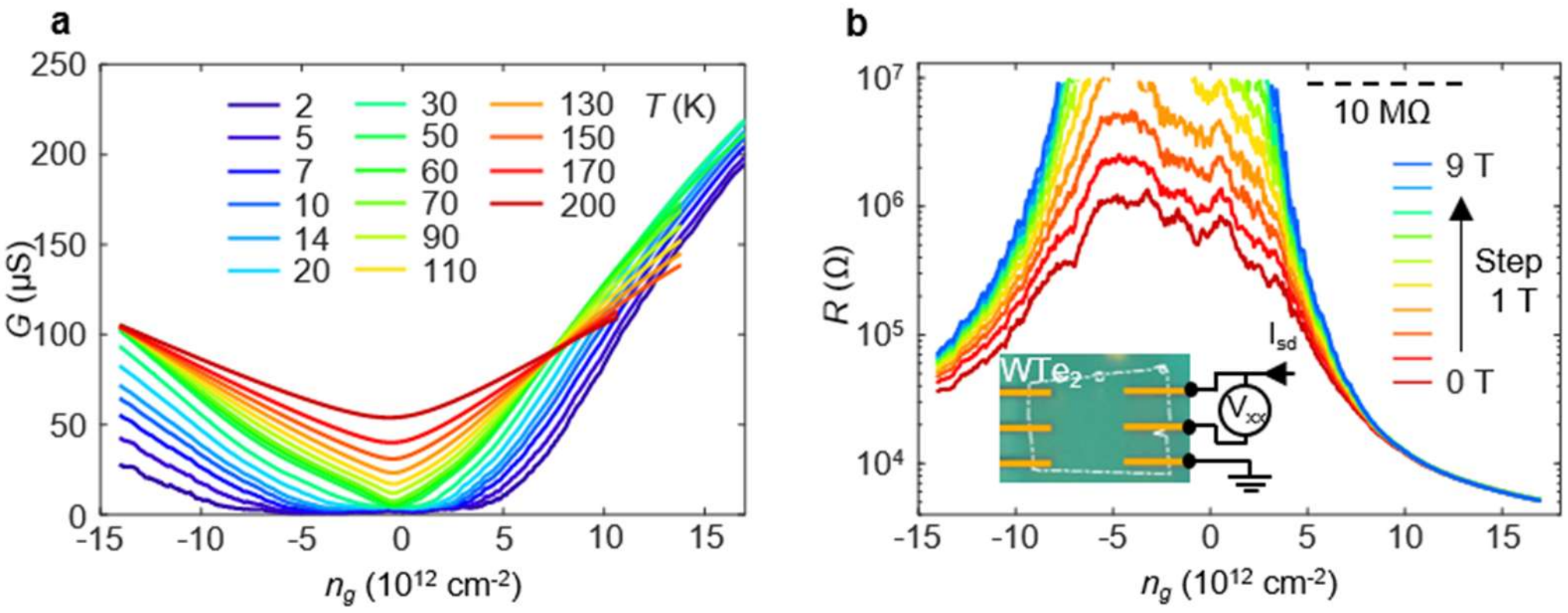

**Extended Data Fig. 2. Electrical transport characterization of device D2.** (**a**) The conductance $G$ of monolayer $WTe_2$ as a function of $n_g$ taken at various $T$ as labeled. (**b**) The resistance ($R\equiv 1/G$) tuned by $n_g$ taken in various B. The measurement geometry is shown as an inset of **b**.

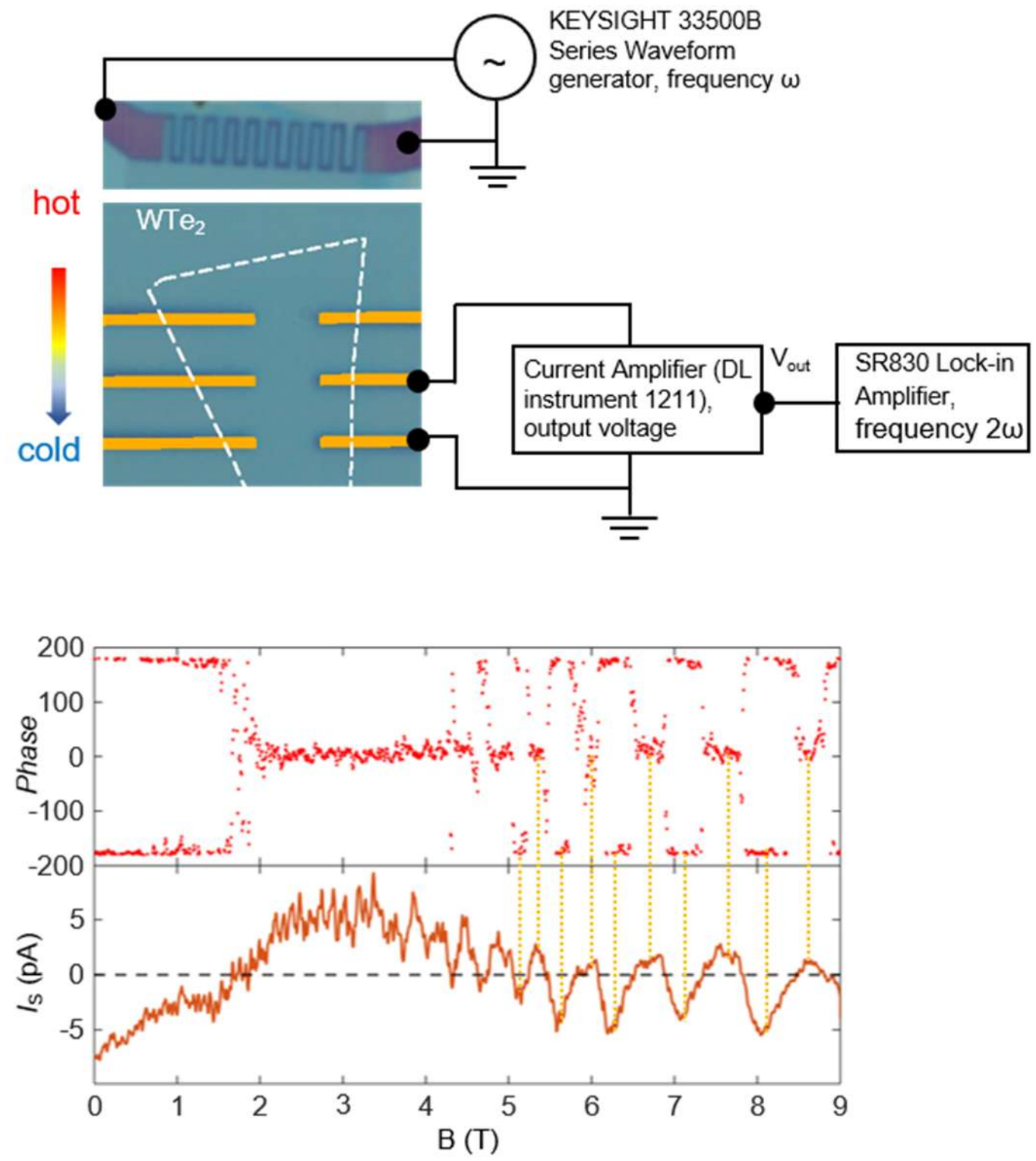

**Extended Data Fig. 3. Measurement configuration (*I*S) and raw data.** Top: detailed measurement setup. Bottom: the magnetic field dependence of signal, including raw data of the output phase (upper) of the lock-in amplifier and the trace of $I_S$ (lower). The data shown corresponds to **Fig. 2d.** All data presented in this work are raw data.

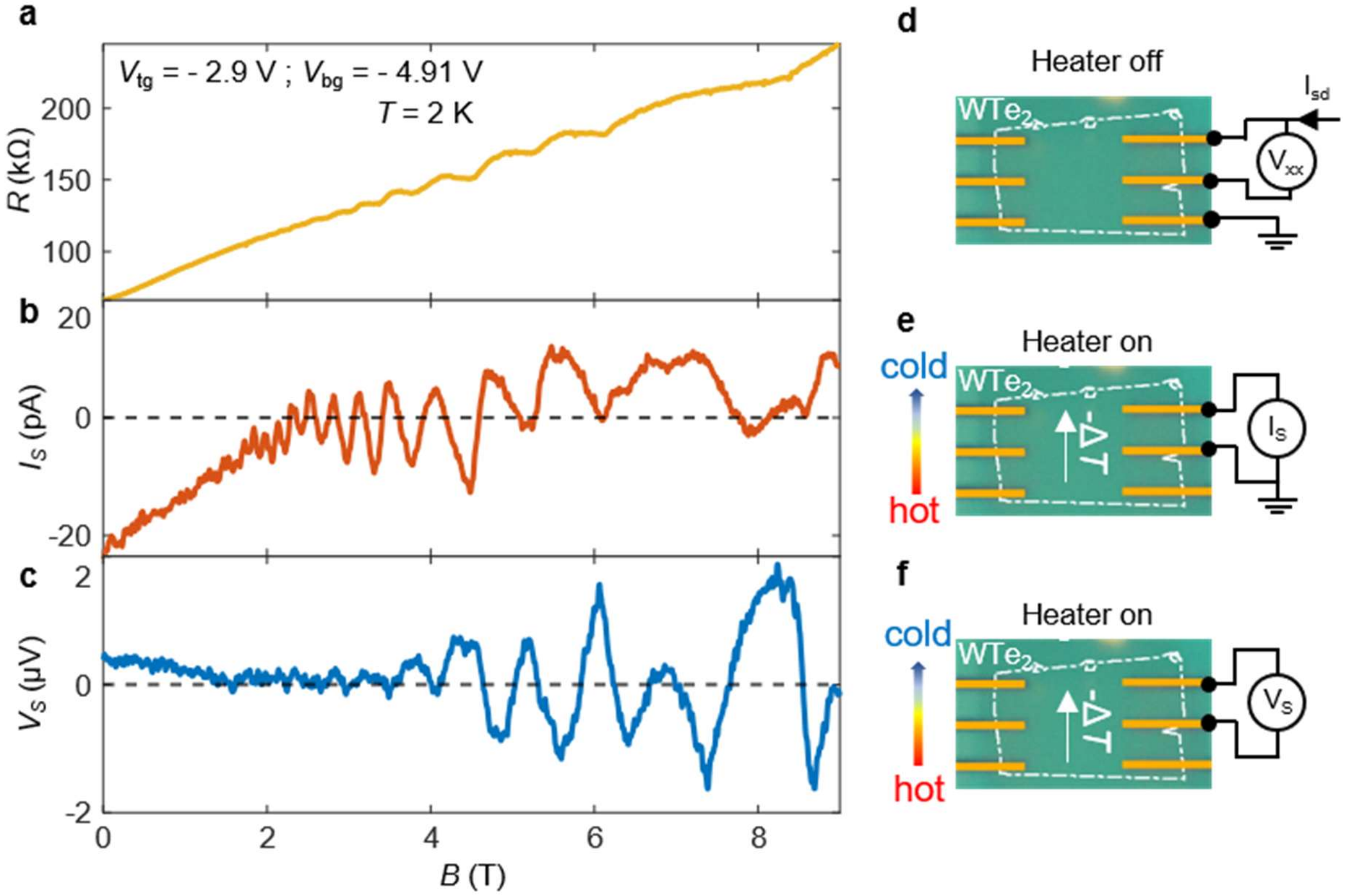

**Extended Data Fig. 4. Sign-change thermoelectric QOs of the hole-doped monolayer $WTe_2$ in device D2.** (**a**) Magnetoresistance of monolayer $WTe_2$ of device D2 taken at $V_{tg}$ = - 2.9 V and $V_{bg}$ = - 4.91 V. The measurement configuration is shown on the right. (**b**) The thermoelectric current response of the same state, with the measurement configuration shown on the right. (**c**) The Seebeck voltage of the same state, with measurement configurations shown on the right. The dashed horizontal lines indicate the zero value.

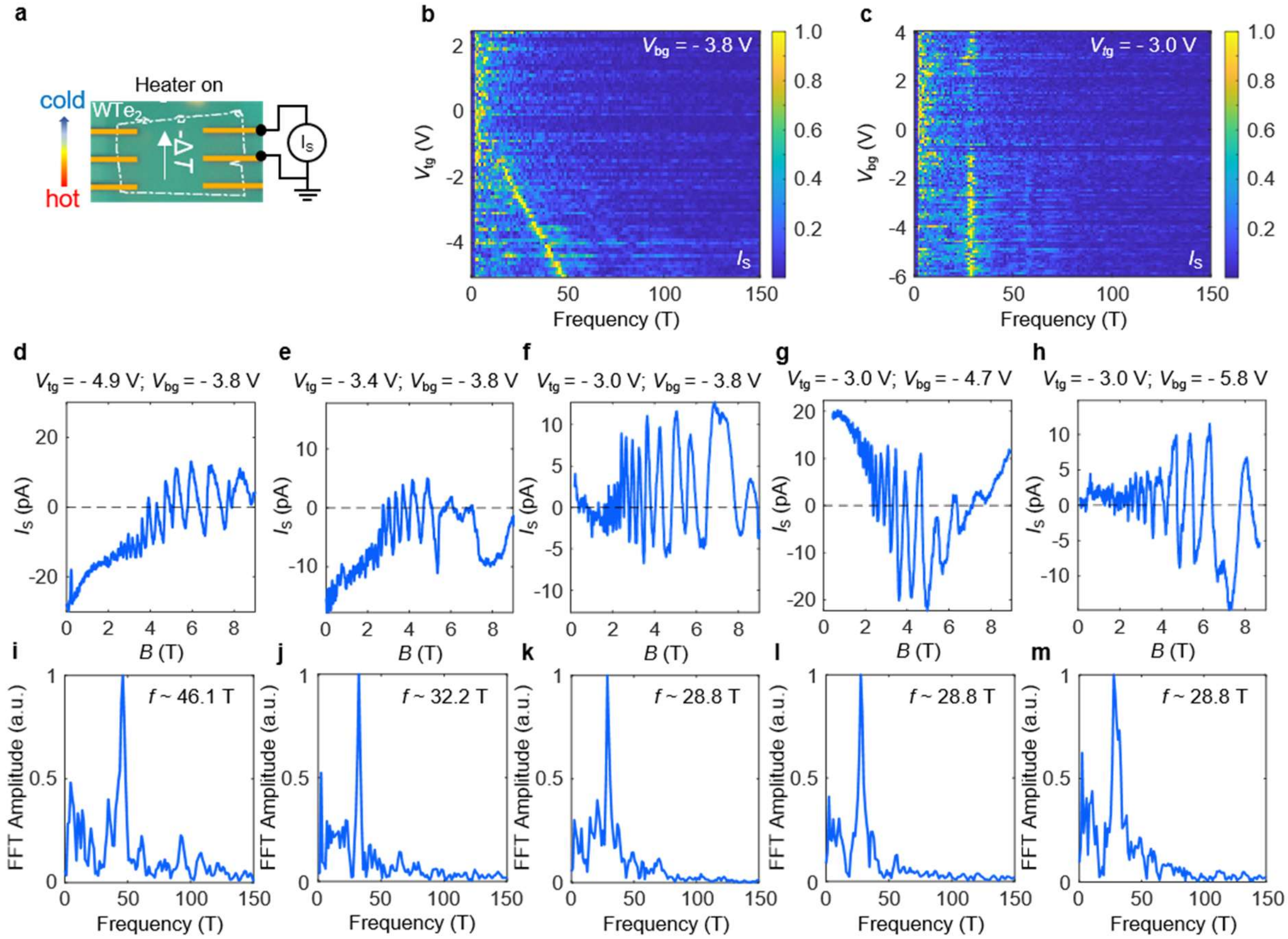

**Extended Data Fig. 5. Gate dependence of thermoelectric QOs in device D2.** (**a**) Measurement configuration for detecting the thermoelectric current ($I_S$). (**b**) A map of normalized FFT amplitude of QOs observed in $I_S$ under varying $V_{tg}$. $V_{bg}$ = -3.8 V. (**c**) A map of normalized FFT amplitude of QOs observed in $I_S$ under varying $V_{bg}$. $V_{tg}$ = -3.0 V. (**d-h**) $I_S$ as a function of $B$ taken under various gate configurations as indicated. Dashed lines indicate the zero value. (**i-m**) Normalized FFT of the QO data shown in (**d**-**h**). The QO frequency ($f$) from FFT analysis is indicated in each panel.

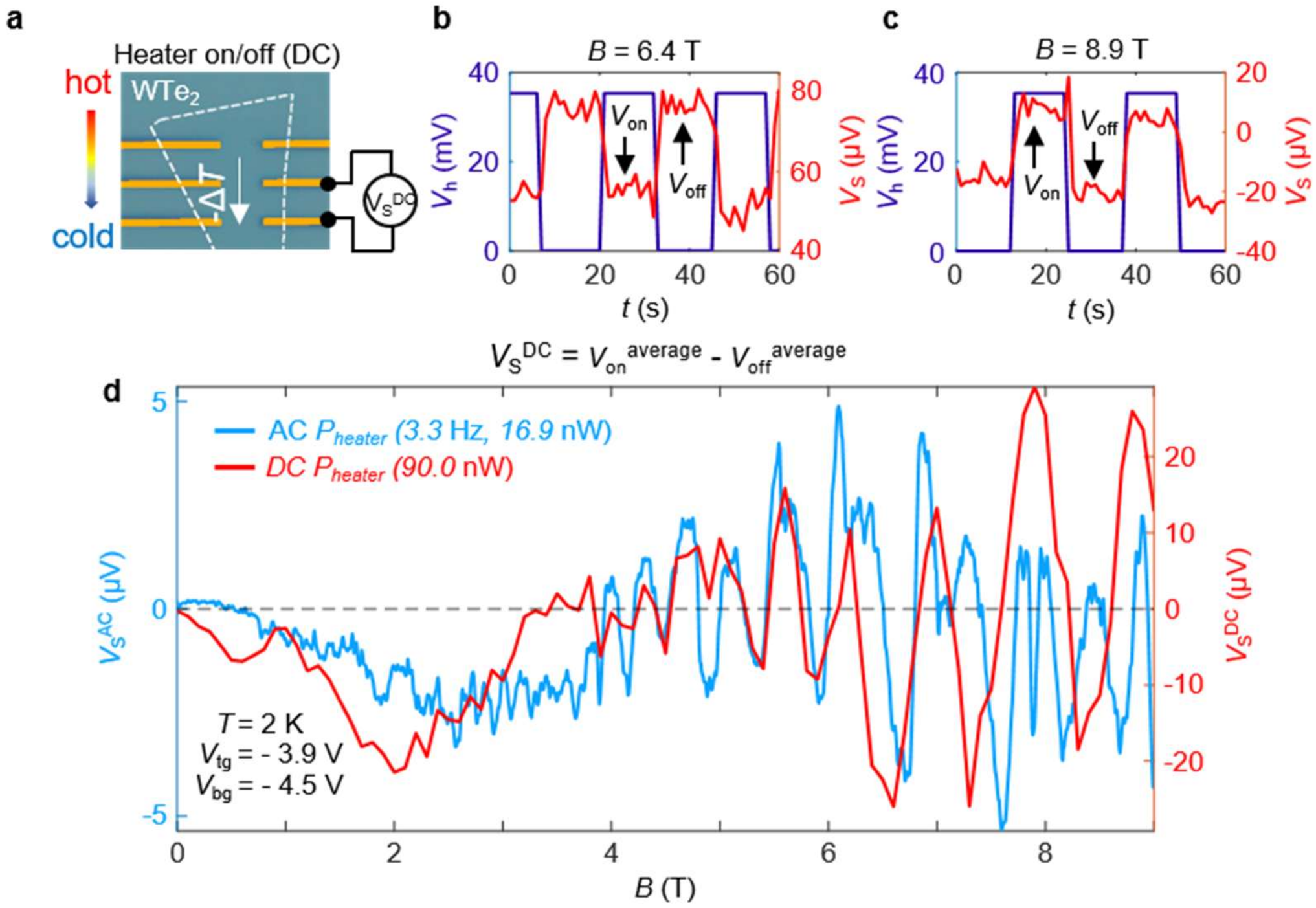

**Extended Data Fig. 6. DC Measurement of QOs in thermopower in device D1.** (**a**) The thermopower measurement configuration with a DC voltage $V_h$ applied to the heater. The temperature gradience when the heater is on is illustrated on the left. (**b**) An on/off train of $V_h$ (blue) applied to the heater, as a function of time, and the corresponding thermoelectric voltage $V_S$ (red) measured through contacts indicated in **a**. B = 6.4 T. When the heater is off, there is background voltage $V_{off}$ developed due to instrumental offset and/or preexisting temperature gradient on the sample etc. In AC measurement, the lock-in technique excludes the detection of this DC background. When the heater is tuned on, a different voltage $V_{on}$ is now measured. The thermoelectric signal due to the applied temperature gradient due to the heater is hence $V_s^{DC} = V_{on} - V_{off}$. Note that both $V_{on}$ and $V_{off}$ are pure DC signals as we parked all voltages at static values. (**c**) The same DC measurements, taken at B = 8.9 T. Note that here $V_{on} > V_{off}$, opposite to that of (**b**), directly proving the sign change of the thermopower signals induced by tuning $B$. (**d**) The comparison of the thermopower measured with DC method source (red) and the low frequency AC lock-in technique (light blue), showing consistent results, including sign changes. The low frequency AC technique provide better sensitivity to signals and faster measurement, and is used for data in main text. All data here were taken at $V_{tg}$ = -3.9 V, $V_{bg}$ = -4.5 V, and a base $T$ of 2 K.

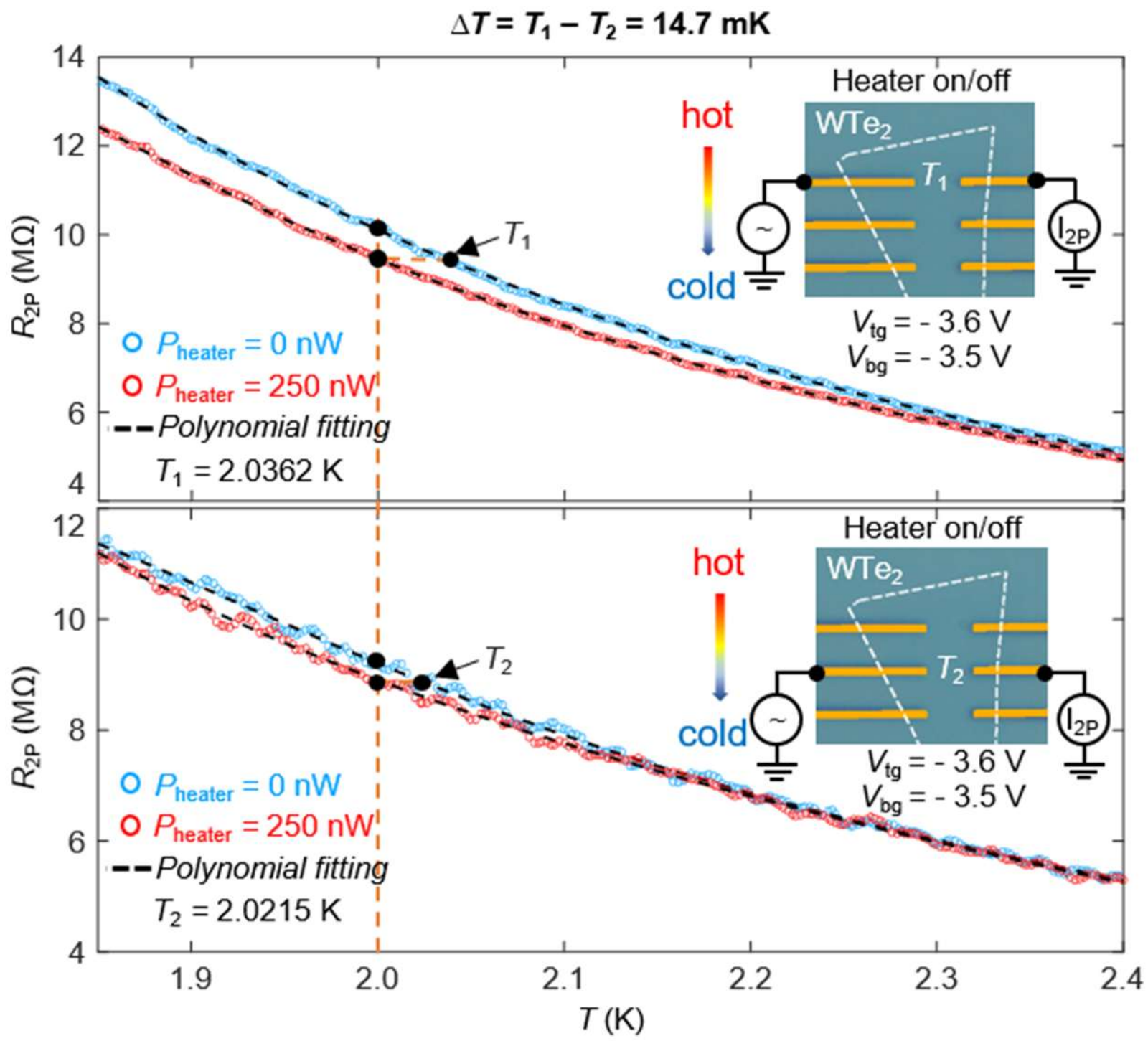

**Extended Data Fig. 7. Temperature gradient calibration.** The plots are temperature-dependent two-probe resistance, with contact configuration illustrated as insets, when the heater is turned off (blue curve) or on (red). The top and bottom panels are data taken at two different local pairs of contacts, as shown. The vertical orange dashed line implies the resistance when sample is at 2 K, when the heater is off (blue data) or on (red data). The decrease of resistance, due to the heater, suggesting an increase of temperature, as expected. The local temperatures when the heater is on can then be extracted by following the horizontal orange dashed lines, labeled as $T_1$ and $T_2$ for the two local contact pairs, respectively. The black dashed lines are polynomial fittings. The temperature difference is estimated by $\Delta T = T_1 - T_2 \sim 14.7$ mK, for the heater power of 250 nW. We use this heater power to provide better reading for the temperature. Given that the longitudinal distance between the contact pairs is ~ 1.5 μm, we then estimate the temperature gradience to be 9.8 mK/μm. In our experiments, only ~ 16.9 nW heater power is used typically, and consequently the temperature gradient is even much smaller.

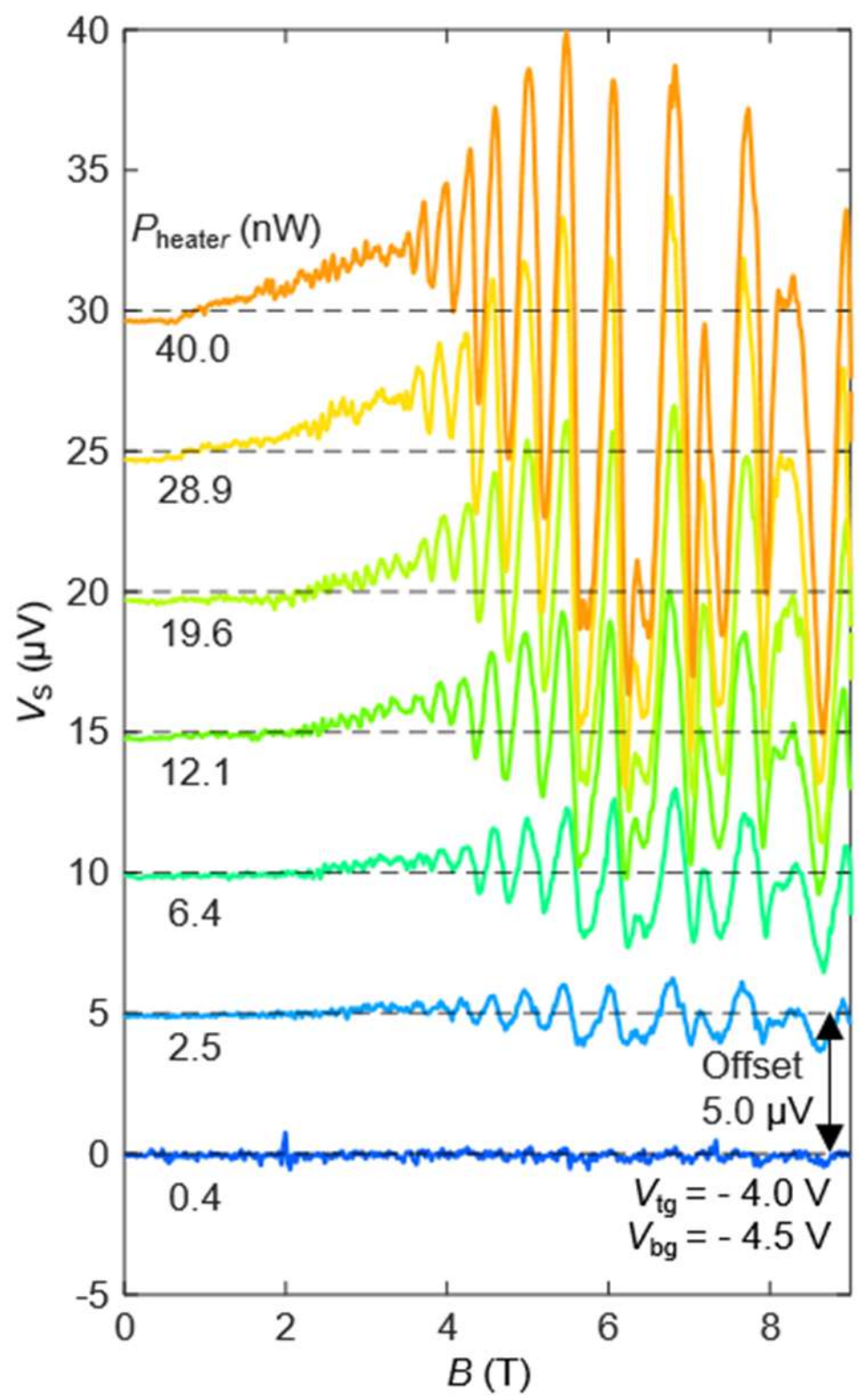

**Extended Data Fig. 8. Heater-power dependence of the QOs.** $V_S$ as a function of $B$, taken at $V_{tg}$ = - 4.0 V and $V_{bg}$ = - 4.5 V, under the application of different heater power ($P_{heater}$) as indicated. Data is taken in device D1. For clarity the curves are offset by 5 μV with respective to each other. The dashed lines indicate the zero value of $V_S$ for each curve.

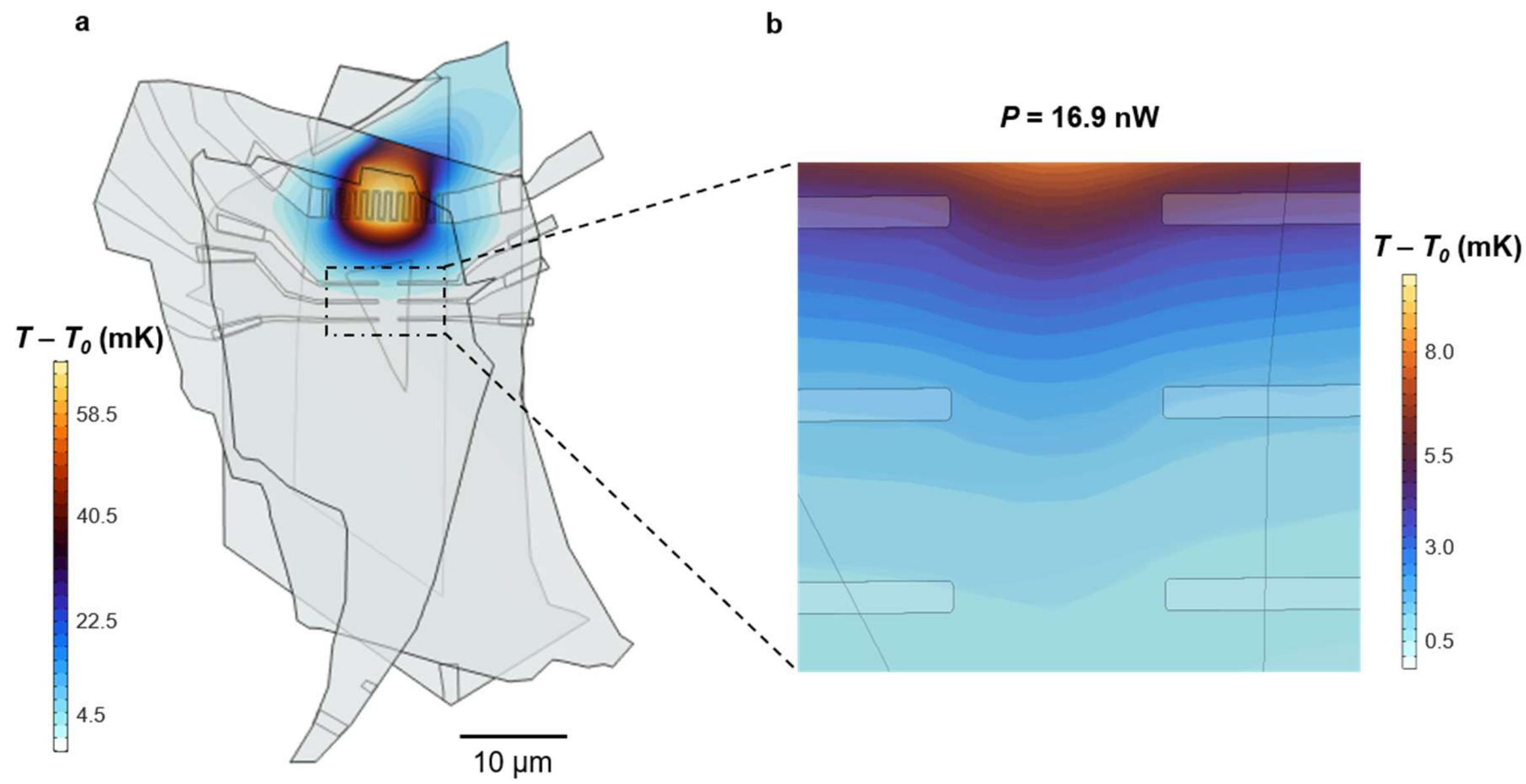

**Extended Data Fig. 9. COMSOL simulation of the temperature gradient of device D1.** (**a**) Schematic of computational model of the full material stack from top view and the simulation result of the temperature distribution under a heater power of 16.9 nW. The zoomed-in distribution near the metal contacts is presented in (**b**). Details of the simulation is described in Methods. $T_0$ = 2 K is the base temperature without heating.

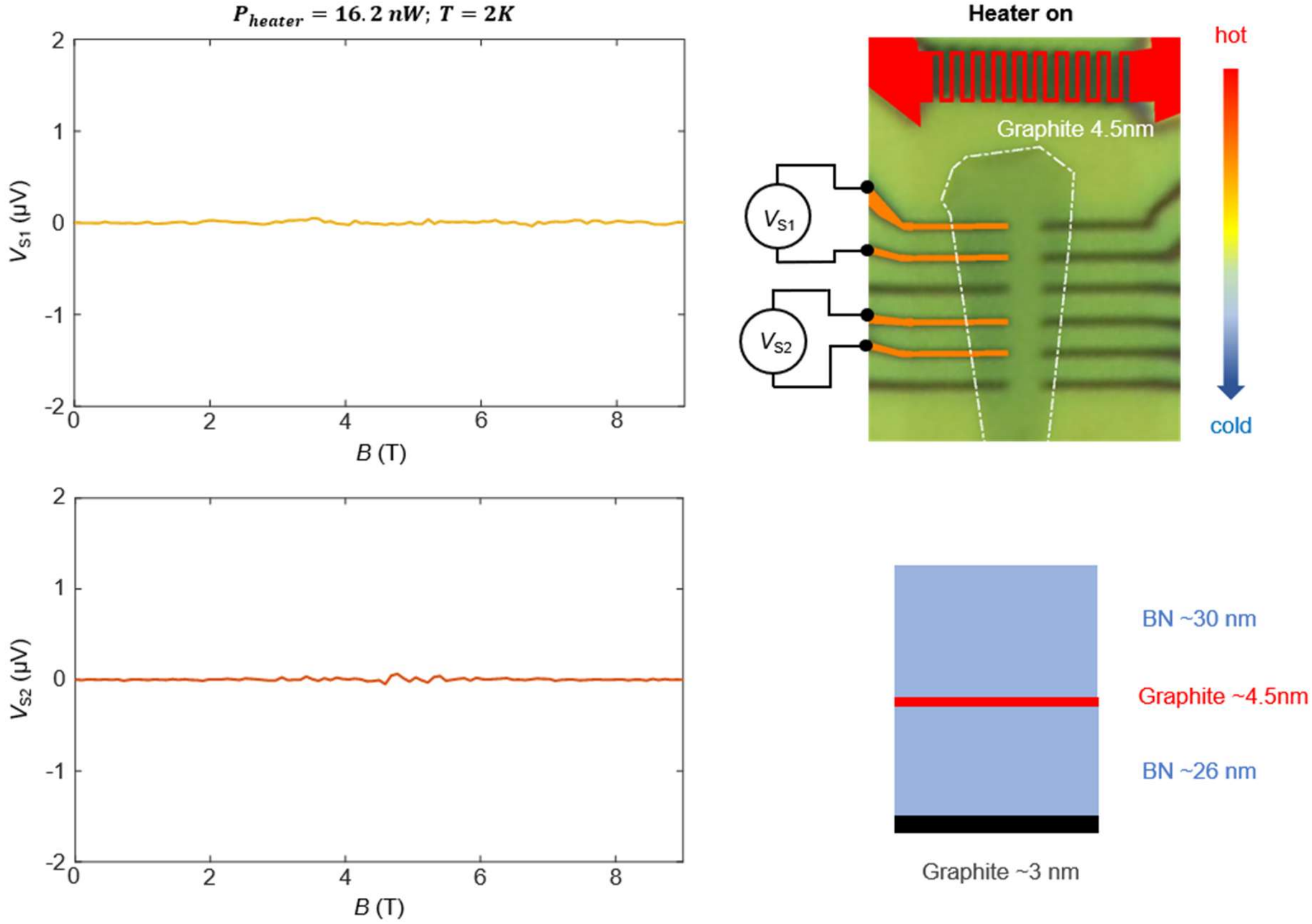

**Extended Data Fig. 10. Absence of graphite thermoelectric voltage signals with a heater power of 16.2 nW, which is used for the $WTe_2$ experiments.** The thermoelectricity measurement of a graphite layer of ~ 4.5 nm thickness (typical for the gate graphite used in our devices) under similar contacts and heater geometry used in the $WTe_2$ experiments. An optical image of the device is shown to the right, with the contact configuration for thermopower measurements illustrated. The cross-section diagram of the device is shown to the bottom right. The measured thermoelectric signals $V_{S1}$ and $V_{S2,}$ as a function of a vertical magnetic field $B$, taken at 2K with a heater power of 16.2 nW, are presented to the left. No detectable signal is resolved, consistent with the expected low thermopower of graphite due to its metallic nature. To resolve the thermopower signal of graphite, one has to apply a substantially larger heater power (i.e., temperature gradient), which is shown in **Extended Data Fig. 11**.

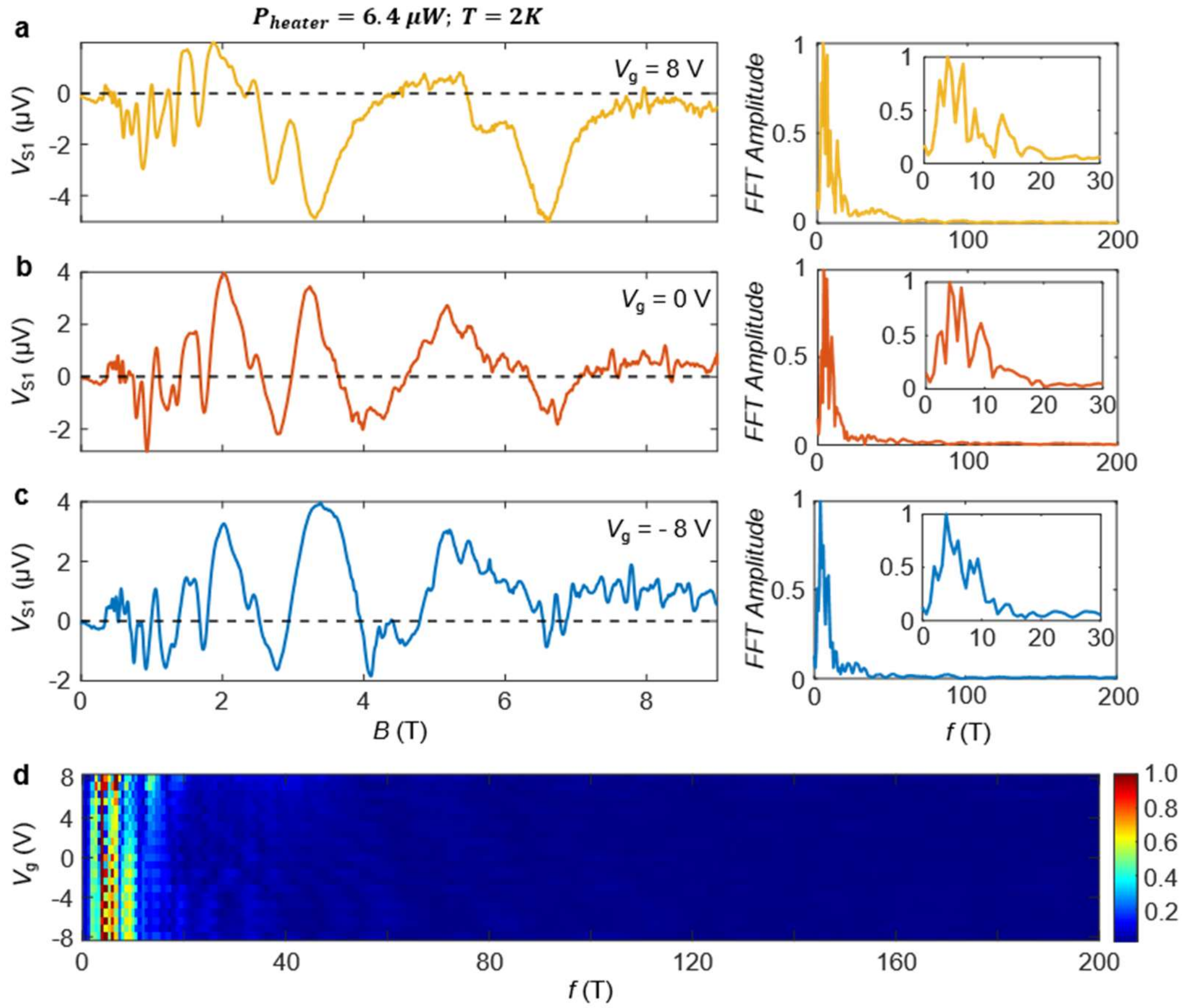

**Extended Data Fig. 11. The Seebeck effect of graphite revealed by a high heater power of 6.4 μW (> 350× higher) and the true QOs of graphite.** The thermopower signal and quantum oscillations under three selected gate voltage applied to a graphite back gate, with (**a**) $V_g = 8$ V (yellow), (**b**) $V_g = 0$ V (red), and (**c**) $V_g = -8$ V(blue). The measurement configuration and device geometry are shown in **Extended Data Fig. 10**. Their corresponding FFT of the QOs is presented to the right for each curve with all peaks located below 20 T. (**d**) A map of normalized FFT amplitude of graphite's QO, showing little gate dependence, consistent with the high carrier concentrations in the graphite. The QOs of the graphite layers are clearly different from that of $WTe_2$, in both its peak locations and the gate dependence (for instance, the QO branch observed in $WTe_2$ reaches a frequency of > 50 T and can be fully depleted by a gate voltage).